\documentclass[useAMS]{mn2e}
\usepackage{psfig,epsfig}
\usepackage{graphicx}
\usepackage{amssymb}
\usepackage{epstopdf}

\begin{document}
\def\simlt{\mathrel{\rlap{\lower 3pt\hbox{$\sim$}}
        \raise 2.0pt\hbox{$<$}}}
\def\simgt{\mathrel{\rlap{\lower 3pt\hbox{$\sim$}}
        \raise 2.0pt\hbox{$>$}}}

\title[Cosmic dichotomy in the hosts of rapidly star-forming systems at low and high redshifts]{Cosmic dichotomy in the hosts of rapidly star-forming systems at low and high redshifts}

\author[Manuela Magliocchetti et al.]
{\parbox[t]\textwidth{M. Magliocchetti$^{1}$,
A. Lapi$^{2,3}$, M. Negrello$^{4}$
G.De Zotti$^{4,3}$, L.Danese$^{3,4}$}\\
 \\
{\tt $^1$ INAF-IAPS, Via Fosso del Cavaliere 100, 00133 Roma,
  Italy}\\
{\tt $^2$ Dip. Fisica, Universita' 'Tor Vergata', Via Ricerca Scientifica 1, 00133, Roma Italy}\\
{\tt $^3$ SISSA-ISAS, Via Bonomea, 265, 34136 Trieste, Italy}\\
{\tt $^4$ INAF-Osservatorio Astronomico di Padova, Vicolo Osservatorio 5, 35122 Padova, Italy} }
\maketitle
\begin{abstract}
This paper presents a compilation of clustering results taken from the literature for galaxies with highly enhanced  (SFR $\simeq [30-10^3]$ $M_\odot$/yr) star formation activity observed in the redshift range $z=[0-3]$.
We show that, irrespective of the selection technique and only very mildly depending on the star forming rate, the clustering lengths of  these objects present a sharp increase of about a factor 3 between $z \sim 1$ and $z \sim 2$, going from values of $\sim 5$ Mpc to about $15$ Mpc and higher. This behaviour is reflected in the trend of the masses of the dark matter hosts of star-forming galaxies which increase from $\sim 10^{11.5}$ $M_\odot$ to $\sim 10^{13.5}$ $M_\odot$ between $z\sim 1$ and $z \sim 2$. Our analysis shows that galaxies which actively form stars at high redshifts are not the same population of sources we observe in the more local universe. In fact, vigorous star formation in the early universe is hosted by very massive structures, while for $z\simlt 1$ a comparable activity  is encountered in much smaller systems, consistent with the down-sizing scenario.
The available clustering data can hardly be reconciled with merging as the main trigger for intense star formation activity at high redshifts.
 We further argue that, after a characteristic time-scale of $\sim 1$~Gyr, massive star-forming galaxies at $z\simgt 2$ evolve into $z\simlt 1.5$ passive galaxies with large ($M_*\simeq [10^{11} - 10^{12}] M_\odot$) stellar masses.
 \end{abstract}

\section{Introduction}
The study of the population of star-forming galaxies has undergone a dramatic acceleration thanks to the advent of the SCUBA submillimetre continuum array receiver and of the {\it Spitzer} and {\it Herschel} satellites which for the first time allowed the investigation of the evolution of the majority of such sources -- those whose activity is enshrouded by dust -- up to redshifts $\simgt 3$. Star formation rates (SFRs) at high redshifts were found to reach spectacular levels ($\simgt 10^3$ $M\odot$/yr, e.g. Yan et al. 2005; Casey et al. 2012). 

The interpretation of these objects is  however still controversial. On the other hand, a definite assessment on their nature has become compelling, especially at a time when the astronomical community is witnessing the birth of two amazing programs aimed at mapping the distribution of star forming galaxies up to very large cosmological scales and early epochs: Euclid and SKA.

Three main scenarios for the early evolution of galaxies can be found in the literature: the so-called merger driven evolution model, the star formation fueled by cold flows model and the self-regulated by baryon processes model. 

The merger-driven evolution model addresses the large enhancement in the star formation activity observed at $z\sim 2$ as due to gas-rich major merging events. Two different scenarios have been proposed: the first one (Baugh et al. 2005) requires modest-sized merger-induced starbursts whose bolometric luminosity is greatly enhanced by a top-heavy initial mass function, while the second one (e.g. Narayanan et al. 2009)  relies on major mergers of large and gas-rich galaxies as the trigger for the starburst phase.
Amongst the many predictions from these two scenarios are small masses for the $z\sim 2$ active galaxies in the first case (Almeida, Baugh \& Lacey, 2011; Kim et al. 202) and large masses and a very short duration of the starburst phase ($\rm{T_{SF}}\sim 0.1$~Gyr) in the second case (Narayanan et al. 2009).

The star formation fueled by cold flows scenario introduced by Fardal et al. (2007) (see also Dekel, Sari \& Ceverino  2009) instead explains the high redshift starburst phenomenon as the product of smooth and steady accretion of gas (and minor mergers) onto massive galaxies. Thanks to the dense intergalactic medium (IGM) and the short cooling times  expected at these high redshifts (see e.g. Dekel \& Birnboim 2006), galaxies will have large accretion rates and therefore will form stars at high rates on relatively long timescales. This model therefore predicts large masses for $z\sim 2$ star forming galaxies, and relatively long timescales  which then convert in a  long duty cycle ($\sim 50$ per cent, i.e. one of such galaxies out of two will be observed in the active star forming phase, Dav\'e et al. 2010).

The third model is the so-called galaxy formation by self-regulated baryon processes. This was first introduced by Granato et al. (2004) and subsequently implemented by Lapi et al. (2006) and Cai et al. (2013). In this model, which also predicts relatively large masses and  long timescales for the high-$z$ starburst phenomenon,  star formation is triggered by the fast collapse of the dark matter halo and is controlled by self-regulated baryonic processes, such as the rapid cooling of the available gas and energy feedbacks from supernovae (SN) and active galactic nuclei (AGN). AGN feedback is particularly relevant for the most massive galaxies and is responsible for the shorter duration ($\sim 0.5-1$ Gyr) of their active star-forming phase. 
In less massive galaxies, the star formation rate is mostly regulated by supernovae feedbacks and continues for a few Gyrs.

As already discussed, the above models predict very different behaviours and physical properties for the galaxy population responsible for the intense star forming activity witnessed at redshifts $\sim 2$. This is particularly true for what concerns galaxy and halo masses and duration of the starburst phase. They also make different predictions on which objects these galaxies will end up into in the more local universe.  This is why clustering measurements at different redshifts are a very important tool to discriminate amongst different scenarios, and can provide the ultimate answer on the nature of such sources.

As a  first step in this direction we have then collected from the literature and subsequently analyzed clustering  results derived for very active -- $SFR\simeq [20-10^3]$ $M_\odot$/yr -- star-forming galaxies in the redshift range $z=[0-3]$. Galaxies observed  in different wavebands (from the UV to radio/HI) were grouped into classes of sources selected at approximately the same rest-frame frequency, and with comparable luminosities, so to overcome any possible bias stemming from selection effects.
Furthermore, clustering results were homogenized to allow for a direct comparison between the various estimates. 

The layout of the paper is as follows: \S 2 introduces the different samples considered in our analysis and provides some estimates for their bolometric luminosities and star formation rates, while \S 3 presents a compilation of clustering lengths $r_0$ as taken from the original works and homogenized to take into account variations in the cosmological parameters and in the slopes of the estimated two-point correlation function amongst the different samples. \S 4 provides our results for the redshift evolution of the halo masses of 
star-forming galaxies between $z\sim 3$ and $z\sim 0$, while \S 5 discusses their cosmological evolution. \S 6 summarizes our conclusions.

Throughout this paper we assume a $\Lambda$CDM cosmology with $H_0=70 \: \rm km\:s^{-1}\: Mpc^{-1}$ ($h=0.7$), $\Omega_0=0.3$,  $\Omega_\Lambda=0.7$ and $\sigma^m_8=0.8$.

\section{The Data}
In order to investigate the properties of star-forming systems at the different cosmological epochs, we have collected from the literature clustering results which refer to galaxies with highly enhanced  star formation activity (i.e. galaxies at the high end of the SFR distribution), selected in different wavebands: UV, sub-mm, mid-infrared (IR), far-IR,  optical/near-IR (BzK) and radio/HI. We further required  the redshift distribution of each sample  to be known with good accuracy (spectroscopic and/or good quality photometric redshifts). Those samples which include data at both low and high redshifts have been divided into selection classes as follows:\\
(i) Far-IR selection (group A)\\
-- Low redshift: IRAS-QDOT galaxies (Saunders, Rowan-Robinson \& Lawrence 1992).\\
-- Intermediate redshift: $z\simlt 1.2$ Herschel galaxies respectively brighter than 8 and 5 mJy selected at 100$\mu$m  by the PEP survey (Lutz et al. 2011) in the COSMOS and Extended Groth Strip (EGS) fields (Magliocchetti et al. 2013).\\
-- High redshift: $z=[1.7-2.6]$ galaxies from the PEP survey of the GOODS-S field, selected at both 100$\mu$m and 160$\mu$m (Magliocchetti et al. 2011).\\
(ii) Sub-mm selection (group B)\\
-- High redshift: $1\simlt z\simlt 3$, 870$\mu$m-selected LABOCA sources in the Extended Chandra Deep Field South (ECDFS;  Hickox et al. 2012).\\
-- Low redshift: 250$\mu$m-selected galaxies from the Herschel-ATLAS Science Demonstration Phase field (van Kampen et al. 2012).\\
(iii) UV selection (group C)\\
 -- Low redshift:  $0.6\simlt z\simlt 1.2$ CFHTLS galaxies selected in the $u^\prime$ band (Heinis et al. 2007).\\
-- High redshift: Hubble Deep Field North (HDFN) galaxies in the redshift range $z=[2.4-3.2]$ (Magliocchetti \& Maddox 1999).

A point which is important to keep in mind in the following analysis is that, within the same selection class, galaxies at the different redshifts  roughly probe the same rest-frame wavelengths. In fact, the low-$z$,
{\bf $u^\prime$} selection mirrors the $z\simgt 2$,  I-band selection of the HDFN. In the same way, $z\sim 2$, $870\,\mu$m-selected LABOCA galaxies can be considered the high-redshift counterparts of the local H-ATLAS sources observed at $250\,\mu$m, and the IRAS $60\,\mu$m-selection at $z\sim 0$ corresponds to that of Herschel galaxies  observed at $\sim100\,\mu$m at $z\sim 1$ and at $\sim 160\,\mu$m at $z\sim 2$ (cfr. Magliocchetti et al. 2013). This minimizes selection biases. 

Furthermore, in order to get a comprehensive and panchromatic view on their properties, we  also add to the aforementioned classes star-forming galaxies selected with different techniques at various redshifts. Locally, we consider the results from 40\% of the Arecibo Legacy Fast ALFA (ALFALFA; Martin et al. 2012) survey which blindly searches the sky for local ($z\simlt 0.06$) HI emitters (group D), while at $z\sim 2$  we also include star-forming galaxies selected on the basis of their mid-IR emission (group E).  In this case, data come from
1) the work of Brodwin et al. (2008) on galaxies selected at 24$\,\mu$m in the Bootes Field with mid-IR-to-optical (R-band) flux density ratios $F_ {24\,\mu\rm m}/F_R>10^3$; 2) the work of Magliocchetti et al. (2008) on galaxies selected in the UKIDSS Ultra Deep Survey (UDS) with $F_ {24\,\mu\rm m}\ge 400\,\mu$Jy; 3) the work by Starikova et al. (2012), which considers $24\,\mu$m-selected galaxies with $F_ {24\,\mu\rm m}\ge 310\, \mu$Jy in the Lockman Field.

 Note that the mid-IR selection only includes galaxies  with $z\simgt 1.5$. This is because the 24$\,\mu$m selection at lower redshifts includes a non-negligible fraction of AGN-powered galaxies (e.g. Gruppioni et al. 2008) which would constitute a 'contaminant' to our sample of purely star-forming galaxies. We also note that, at variance with the other works, those of Brodwin et al. (2008) and Starikova et al. (2012) base their results on models for the source redshift distribution obtained from extrapolation from other datasets. This might imply larger uncertainties in their results and most likely an underestimate of the true clustering signal due to its dilution by interlopers mistakenly assumed to belong to a chosen redshift range.  Lastly, we  also consider the clustering results of Lin et al. (2012) 
who use the BzK technique to select star-forming galaxies in the GOODS North field  (group F).
In order to make the Lin et al. (2012) 
$z\sim 2$ data comparable with those considered in this work, we only include galaxies which present the highest-estimated star-formation rates, i.e. respectively  SFRs $\ge 30$ M$_\odot$/yr,  SFRs $\ge 60$ M$_\odot$/yr and SFRs $\ge 100$ M$_\odot$/yr. 

Table 1 summarizes the data sets considered in our analysis. The minimum IR (8--$1000\,\mu$m) luminosities  averaged over the redshift distributions, $\langle L_{\rm IR,min}\rangle$, of galaxies selected in the mid-IR to sub-mm wavelength range were computed as
\begin{eqnarray}
\langle L_{\rm IR, min}\rangle =\frac{\int^{z_{\rm max}}_{z_{\rm min}}  L'_{\rm min}(z) N(z)\; dz }{\int^{z_{\rm max}}_{z_{\rm min}} N(z)\; dz},
\end{eqnarray}
with $L'_{\rm min}(z)=\int l_{\rm min}(\lambda^*,z)\; f(\lambda) \; d\lambda$, where $l_{\rm min}(\lambda^*,z)=4\pi S_{\rm min} d_L^2/K(\lambda^*,z)$ is the minimum monochromatic luminosity at $\lambda^*$ corresponding to $S_{\rm min}$, $d_L$ is the luminosity distance and  the K-correction is  expressed  as $K(\lambda^*,z)=(1+z)f(\lambda^*/(1+z))/f(\lambda^*)$. Based on the results by Gruppioni et al. (2010), the normalized emission spectrum $f$ was taken to be M82-like at low-to-intermediate redshifts and Arp220-like at $z\sim 2$. However, we stress that the results do not greatly vary if we assume the same spectrum for all the objects under exam. The redshift distributions $N(z)$ in eq.~(1) were taken from the corresponding papers. Star formation rates were then derived using the standard relation (Kennicutt 1998): $\rm SFR [M_\odot/yr]=1.8\times 10^{-10} L_{\rm IR}/L_\odot$.  

 As already anticipated, all the sources included in our analysis present an intense star forming activity, with SFRs ranging from $\sim 20-30$ $M_\odot$/yr up to values of the order of a few $\times 10^3 M_\odot$/yr, with the possible exception of the Martin et al. (2012) sample, which probably includes less active objects. 
We also note that all the works presented here are truly SFR-selected samples, either because most of the stellar light in these rapidly star-forming objects is dominated by a young population and so the observed luminosities depend weakly on the stellar mass (as is the case of galaxies selected in the mid/far-IR), because they are detected purely on emission from gas rather than stars (as for HI emitters), or because the stellar mass limits are always sufficiently deep that all galaxies above the SFR limits are included in each sample (as for Magliocchetti \& Maddox 1999 and Lin et al. 2012).

\begin{table*}
\begin{center}
\caption{Overview of the properties of the star-forming sources considered in our analysis.  The columns refer to:  (1)  selection criterion, (2) observed field, (3) average redshift, (4) minimum flux (in mJy, rows 1-7 and 12-15) or limiting magnitude (rows  8-10),  (5) Log$(\langle L_{\rm min} \rangle)$ in solar units, (6) minimum SFR in ${\rm M}_\odot$/yr, (7) clustering length, in Mpc, for $\gamma=1.8$, (8) Log of the minimum mass in solar units and references to the various works (column 9). Samples observed at approximately  the same rest-frame wavelength have been labeled with the same capital letter (A for the far-IR selection, etc.; see text for details).  Minimum SFRs for UV-selected galaxies are bracketed by the two values respectively obtained directly from UV luminosities (lower limit) and via integration of the full SED (upper limit).}
\begin{tabular}{lllllllll}
Selection  & Field&$\langle z \rangle $& S$_{\rm min}$& Log$(\langle L_{\rm min} \rangle)$ &$\langle$SFR$_{\rm min}\rangle$ &$r_0$& Log$(\hbox{M}_{\rm min})$& Reference   \\
\hline
\hline
A-IRAS-[$60\mu\rm m$] & All sky&$\sim$0.02 &600 &$11.0\pm 0.4$& $18^{+27}_{-11} $ &$5.4^{+0.2 }_{-0.2}$&$11.4^{+0.2}_{-0.2}$& {\tiny Saunders et al. 1992} \\
A-Herschel-[$100\mu\rm m$] & EGS& $0.68\pm 0.39$& 5 &  $11.6\pm 0.5$&$72^{+154}_{-50} $&$5.0^{+2.2 }_{-3.3} $&$11.9^{+0.5 }_{-1.1}$ & {\tiny Magliocchetti et al. 2013} \\
A-Herschel-[$100\mu\rm m$] &COSMOS& $0.56\pm 0.36$&8 & $11.6\pm0.6$&$72^{+213}_{-54} $& $4.1^{+0.8 }_{-1.0} $& $11.1^{+0.4 }_{-0.7}$& {\tiny Magliocchetti et al. 2013} \\
A-Herschel-[$160\mu\rm m$]& GOODS-S& $2.1\pm 0.3$& 5 &  $12.1\pm 0.2$&$226^{+133}_{-83} $&$17.4^{+2.8 }_{-3.1} $&$13.7^{+0.3 }_{-0.4}$& {\tiny Magliocchetti et al. 2011} \\
A-Herschel-[$100\mu\rm m$]& GOODS-S& $2.1\pm 0.2$& 2 &  $12.3\pm 0.2$&$372^{+197}_{-145} $&$19.0^{+2.6 }_{-2.9} $&$13.8^{+0.2 }_{-0.3}$& {\tiny Magliocchetti et al. 2011} \\
B-LABOCA[$870\mu\rm m$]&ECDF& $\sim 2.1$& 4.5&  $12.71\pm 0.02$&$923^{+43}_{-42} $&$11.0^{+2.6}_{-3.3} $&$13.0^{+0.3 }_{-0.5}$& {\tiny Hickox et al. 2012} \\
B-Herschel-[$250\mu\rm m$] &SDP+GAMA&$\sim 0.25$& 33&  $11.5\pm 0.2$&$57^{+33}_{-21} $&$5.6^{+1.1}_{-1.1} $&$11.9^{+0.4}_{-0.7}$& {\tiny van Kampen et al. 2012} \\
C-AB(8140) &HDF-N&$2.6\pm 0.2$& 28 & -- &$[30-90]$&$18^{+7}_{-7} $&$13.5^{+0.3}_{-0.6}$& {\tiny Magliocchetti \& Maddox 1999} \\
C-AB(8140) &HDF-N&$3.0\pm 0.2$& 28 & -- &$[40-100]$&$17^{+12}_{-12} $&$13.4^{+0.7}_{-1.5}$& {\tiny Magliocchetti \& Maddox 1999}\\
C-u$^\prime$ CFHTLS& $4\times 1$ deg$^2$&$0.94\pm0.16$& 24& --&[14-150]& $4.6^{+0.5}_{-0.5} $&$11.6^{+0.3}_{-0.3}$& {\tiny Heinis et al. 2007}\\
D-HI ALFALFA& 40\% allsky& $\sim 0.02$ &-- & -- & -- & $4.8^{+0.4}_{-0.3} $&$11.0^{+0.4}_{-0.6}$& {\tiny Martin et al. 2012}\\
E-Spitzer-[$24\mu\rm m$]& Bootes& $2.0\pm 0.45$& 0.3&  $12.8\pm 0.3$&$1135^{+1131}_{-565} $&$11.1^{+1.9 }_{-1.2} $&$13.1^{+0.2 }_{-0.2}$&{\tiny Brodwin et al. 2008} \\
E-Spitzer-[$24\mu\rm m$]& Bootes& $2.0\pm 0.45$& 0.6&  $13.1\pm 0.3$&$2266^{^+1130}_{-2255} $&$20.0^{+6.5 }_{-4.1} $&$13.8^{+0.2 }_{-0.3}$& {\tiny Brodwin et al. 2008} \\
E-Spitzer-[$24\mu\rm m$]&Lockman& $1.7\pm 0.63$& 0.31&  $12.8\pm 0.4$&$1135^{+1718}_{-683} $&$11.0^{+0.9}_{-0.9} $&$13.1^{+0.1 }_{-0.1}$& {\tiny Starikova et al. 2012} \\
E-Spitzer-[$24\mu\rm m$]&UDS& $2.1\pm 0.3$& 0.4&  $12.9\pm 0.6$&$1613^{+4079}_{-1253} $&$15.9^{+2.9}_{-3.4} $&$13.3^{+0.2 }_{-0.4}$& {\tiny Magliocchetti et al. 2008} \\
F-BzK &GOODS-N&$\sim 2.21$& -- & -- &100&$20.1^{+5.0}_{-5.0} $&$13.7^{+0.15 }_{-0.15}$&{\tiny Lin et al. 2012} \\
F-BzK &GOODS-N&$\sim 2.27$& -- & -- &60&$17.4^{+4.3}_{-4.3} $&$13.6^{+0.2}_{-0.2}$&{\tiny Lin et al. 2012} \\
F-BzK &GOODS-N&$\sim 2.21$& -- & -- &30&$13.1^{+3.2}_{-3.2} $&$13.2^{+0.2 }_{-0.2}$&{\tiny Lin et al. 2012} \\

\hline
\hline

\end{tabular}
\end{center}
\end{table*}

SFRs for UV-selected sources were instead derived in two different ways: the first one,  $\rm SFR [M_\odot/yr]=1.4\times 10^{-28} L_{\nu}$[ergs/s/Hz] (Kennicutt 1998), relies on the rest-frame UV luminosity and holds  in the range [1500--2800]\,\AA\  for a Salpeter IMF.  The second method instead follows that previously adopted for infrared-selected sources and estimates the SFR from integration of the full spectral energy distribution (Arp220-like SED), normalized to the observed $L_{\nu}$.  Not surprisingly, the two estimates do not agree with each other. In fact, the first method does not take into account dust extinction, proven to be very relevant especially in the redshift range $z\simeq [1-3.]$  (e.g. Burgarella et al. 2013),  and may therefore return values for the SFRs which are strongly underestimated. For instance, Reddy et al. (2012) provide a value of $5.2\pm 0.6$ for the median correction factor needed to recover the true SFR from the UV luminosity of UV-selected galaxies at $z\sim 2$. \\
Since we do not know the dust-extinction correction factor associated to our samples of UV-selected sources, in Table 1 we report both estimates of the SFRs as obtained from the two different methods (lower values for $L_{\rm UV}$-estimated SFRs), even though we expect true values to be closer to those derived from integration of the full SED. 
We note that, even in the most conservative case, we find SFRs $\simgt 20-40$ M$_\odot$/yr.

\section{Variation of the clustering length with cosmic time}

The aforementioned works provide estimates for the correlation length $r_0$, defined as
$\xi(r)=\left(r/r_0\right)^{-\gamma}$,
where $\xi(r)$ is the spatial two-point correlation function. However, only a handful of such works (precisely those of van Kampen et al. 2012, Saunders at al. 1992 and Martin et al. 2012) provide a direct (3D) measurement of $\xi(r)$. All the others, which mainly rely on photometric estimates of the redshifts, measure its projected counterpart, i.e. the angular two-point correlation function $w(\theta)$, in general parameterized as $w(\theta)=A \theta^{1-\gamma}$, and derive the clustering length $r_0$ using the relativistic Limber equation: 
\begin{eqnarray}
w(\theta)=2\:\frac{\int_0^{\infty}\int_0^{\infty}F^{-2}(x)x^4\Phi^2(x)
\xi(r,z)dx\:du}{\left[\int_0^{\infty}F^{-1}(x)x^2\Phi(x)dx\right]^2},
\label{eqn:limber} 
\end{eqnarray}
where $x$ is the comoving coordinate, $F(x)$ gives the correction for
curvature, and the selection function $\Phi(x)$ is related to the redshift distribution $N(z)$ via
\begin{eqnarray}
{\cal N}=\int_0^{\infty}\Phi(x) F^{-1}(x)x^2 dx=\frac{1}{\Omega_s}
\int_0^{\infty
}N(z)dz,
\label{eqn:Ndense} 
\end{eqnarray}
where $\cal N$ is the mean surface density on a surface of solid angle $\Omega_s$. 
The results depend both on the adopted cosmology and on the value of $\gamma$.

In order to correct for the $\gamma$ dependence of the different estimates of $r_0$, following Magliocchetti et al. (2000), we use the rms fluctuations in the galaxy distribution at the scale of 8/h Mpc, $\sigma_8$.
We relate $\sigma_8$ to the quantities $r_0$ and $\gamma$ as
\begin{eqnarray}
\sigma_8=\left[\left(\frac{r_0}{8/h}\right)^{\gamma}c_{\gamma}\right]^{1/2},
\end{eqnarray}
where $c_{\gamma}={72}/[(3-\gamma)(4-\gamma)(6-\gamma)2^\gamma]$ (cfr. Peebles 1980). Once we have $\sigma_8$ estimated from the data, we can invert eq.~(4) to derive the values of $r_0$ corresponding to the chosen $\gamma$. Since most of the works considered in the present analysis use $\gamma=1.8$, this will be our reference value. The $r_0$ estimates which need to be modified are those of Starikova et al. (2012; $\gamma\simeq 1.7$), Brodwin et al. (2008; $\gamma=1.9$), Heinis et al. (2007; $\gamma\simeq 1.7$),  Saunders et al. (1992) and Martin et al. (2012) who respectively derive from their data $\gamma\simeq 1.6$ and $\gamma=1.51$. 

Furthermore, we had to homogenize the results from Lin et al. (2012) (who provide values of $r_0$ for SFR intervals) with the others (provided for SFRs above some threshold). Following Magliocchetti et al. (2013), this was done by considering the expression $\xi(SFR\ge A)=n_{B}^2 \xi_{B} + n_{AB}^2 \xi_{AB}+ 2 \; n_{B} n_{AB} \sqrt{\xi_{B}  \xi_{AB}}$, where $B\equiv SFR_B> A\equiv SFR_A$, and $AB\equiv SFR_{AB}$ indicates the star formation range of values [SFR$_A$-SFR$_B$], while $n_B$ and $n_{AB}$ are the fractions of galaxies respectively with $SFR\ge SFR_{B}$ and $SFR_A\le SFR\le SFR_B$. 

The correlation lengths $r_0$ resulting from the above analysis are reported in Table 1 and plotted in Figure 1 as a function of the average redshift of each considered survey.
It is clear from the plot that, irrespective of the selection method, there is an abrupt jump in the clustering strength of star-forming galaxies between the low and the high-redshift regime. In fact, below $z\simeq 1$ the correlation length is generally in the range 4--6 Mpc, while for $z\simgt 1.7$  $r_0$ leaps up to values between 11 and 20 Mpc. Furthermore, as mentioned earlier, two of the lowest $z\sim 2$ values, precisely those of Starikova et al. (2012; green empty square) and Brodwin et al. (2008; green open circles) might be underestimates of the true quantities.

Could this dichotomy be the result of some selection bias? Hardly so. First of all, we remind that  within each class, galaxies at the different redshifts are observed at the same rest-frame wavelength so that the selection is as homogeneous as possible. Furthermore, as Table 1 shows, the SFRs of high redshift sources span a huge interval, from the $\simgt 30$ $M_\odot$/yr of the Lin et al. (2012) sample, to the few$\times  10^3$ $M_\odot$/yr of {\it Spitzer}-selected sources. There is no obvious dependence  of the clustering strength of these objects on their SFRs, as galaxies with moderate star formation activity, such as those in the sample of Lin et al. (2012) or in the GOODS-S {\it Herschel} dataset of Magliocchetti et al. (2013),  are just as clustered or even more clustered than galaxies with extreme SFRs, such as those selected at 24$\mu$m or at  870$\mu$m. There is a hint for brighter objects within the same selection class to be clustered more strongly than fainter ones (Brodwin et al. 2008; Lin et al. 2012) but, as it is clear from Figure 1, this is a second-order effect which does not affect the present discussion. This result also holds if instead of threshold luminosities one uses luminosity intervals (as it was done in Lin et al. 2012): in fact, also in this case one has that, above SFR$\simgt 30$ $M_\odot$/yr galaxies show strong clustering regardless of  their intrinsic luminosity.

Similarly, the clustering strengths of $z\simlt 1$ star-forming galaxies are strikingly alike irrespective of the selection criterion and of the SFR. Indeed, {\it Herschel}  galaxies with SFR $\simgt 60-100$ $M_\odot$/yr (van Kampen et al. 2012; Magliocchetti et al. 2013) are just as clustered as the less active IRAS or UV-selected sources. Furthermore, the global trend of $r_0$ in the whole $z=[0-3]$ range coincides with that observed within the homogeneous classes of far-IR-selected galaxies and UV-selected galaxies which, as shown in Table 1, are endowed with very similar bolometric luminosities and SFRs at all redshifts. We also note that the Lin et al. (2012) and possibly also the HDFN samples at $z\sim 2$ include galaxies which have luminosities comparable to $z\simlt 1$ sources. 

All the points mentioned above argue against the possibility that the observed trend is caused by either selection or luminosity-dependent effects.

\begin{figure}
\begin{center}
\includegraphics[scale=0.45]{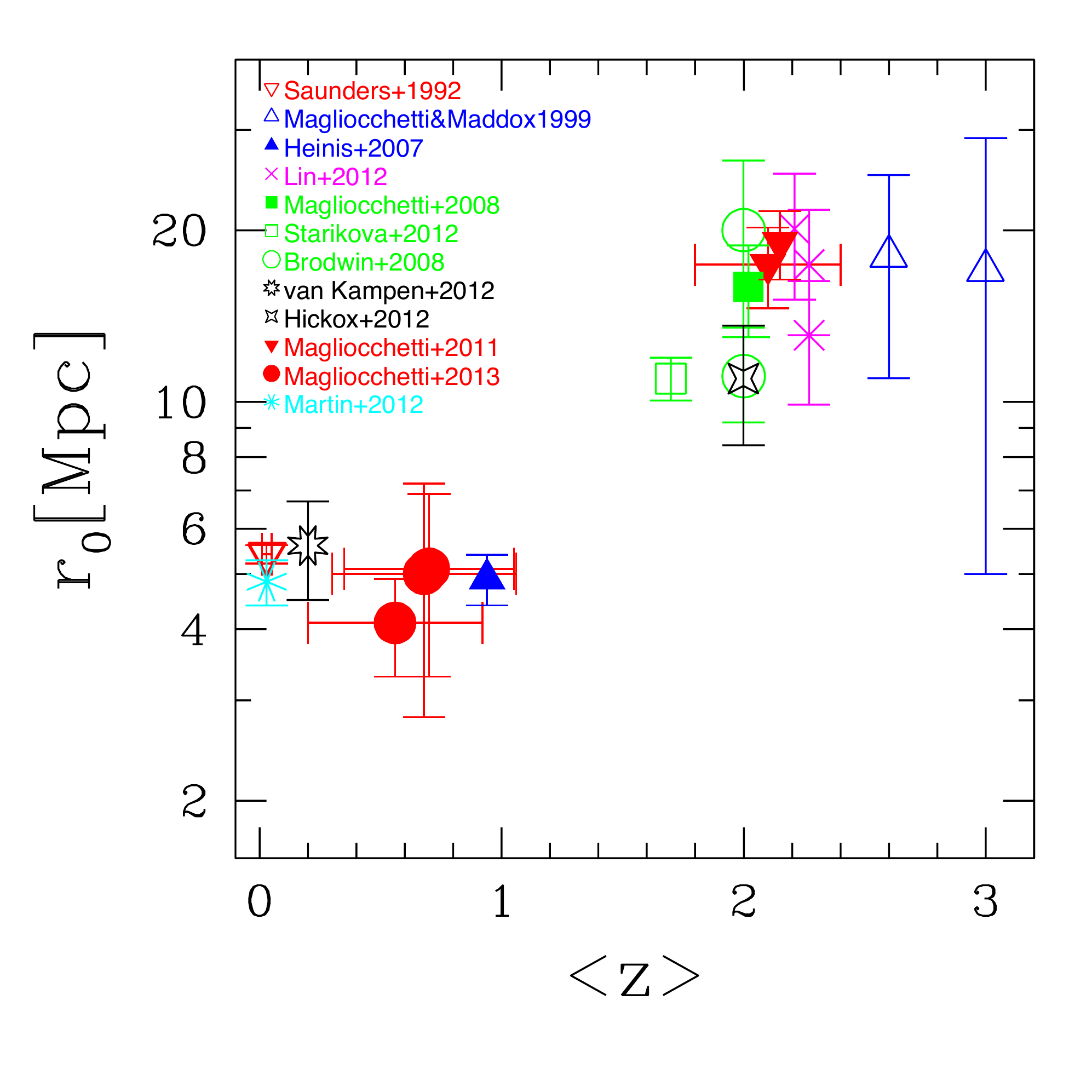}
\caption{Measurements of the comoving clustering length, $r_0$, of actively star-forming galaxies. Different symbols correspond to different samples, while different colours correspond to different selection techniques: red for far-IR selection, blue for UV selection, green for mid-IR selection, black for sub-mm selection, magenta for BzK selection and cyan for radio/HI selection.
\label{fig:r0}}
\end{center}
\end{figure}

\begin{figure}
\begin{center}
\includegraphics[scale=0.45]{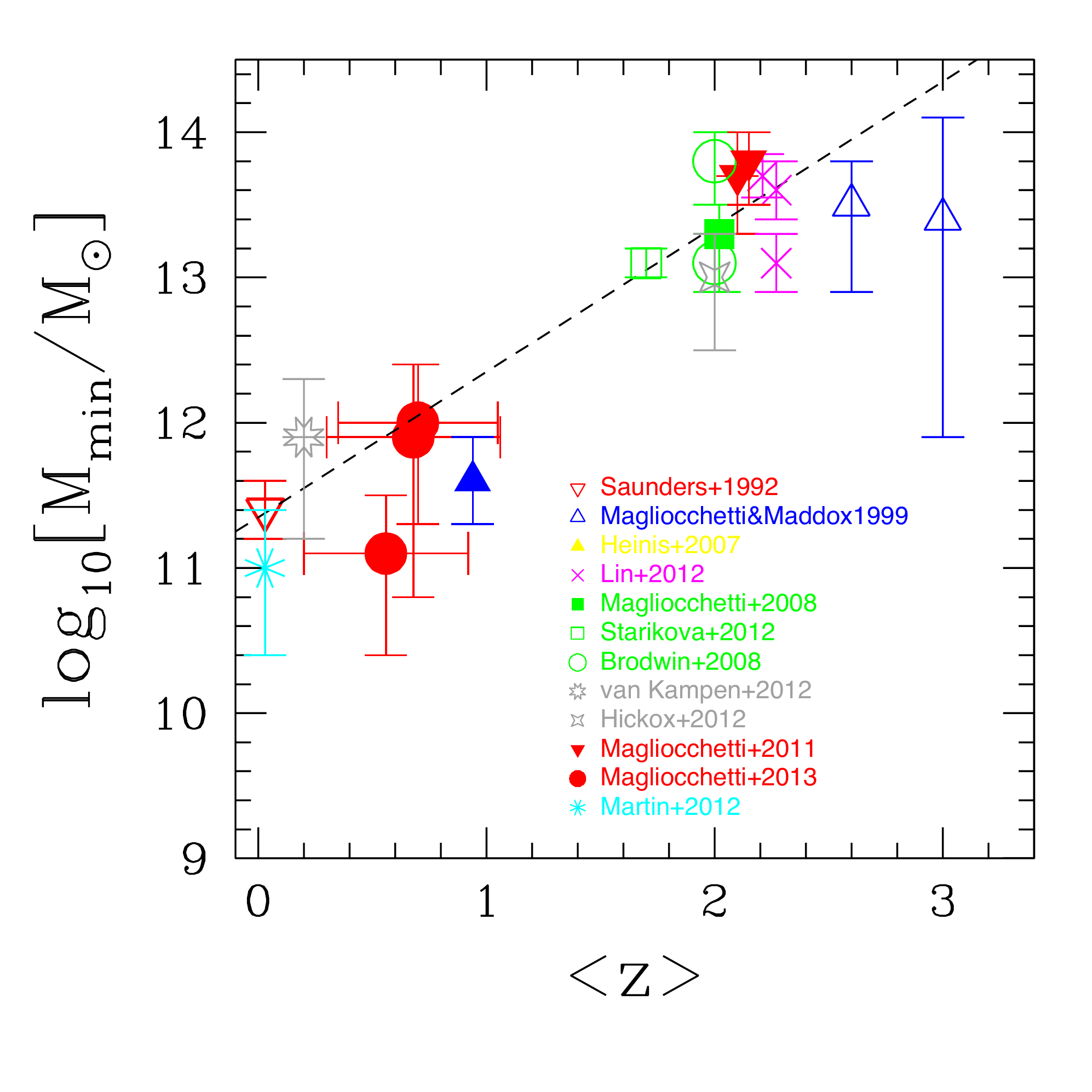}
\caption{Variation with redshift of the minimum halo mass of star-forming galaxies. Symbols and colour coding are the same as in Figure 1. The dashed line represents the best fit to the data obtained for Log[$M_{\rm min} /M_\odot]=\alpha \cdot z+\beta$, with $\alpha= 1$ and $\beta=11.35$.
\label{fig:mass}}
\end{center}
\end{figure}

\section{Variation of the halo mass with cosmic epoch}

The correlation length provides a direct way to estimate the total masses of the haloes where the sources producing the clustering signal reside. However, the relationship between these two quantities is not straightforward, and different approaches have been proposed in the literature leading to different mass estimates based on the same data. For example, the Halo Occupation Model (HOM; see e.g.  Scoccimarro et al. 2001), used by Magliocchetti et al. (2008), yields values of $M_{\rm min}$ which are similar to those obtained from N-body simulations (method used by Starikova et al. 2012 and Brodwin et al. 2008), but systematically lower than those given by the halo bias formalism of Mo \& White (1996) and Sheth \& Tormen (1999), adopted by Magliocchetti et al. (2011; 2013), Hikcox et al. (2012), Lin et al. (2012) and Heinis et al. (2007). 

In fact, as a general rule, within the HOM (which provides analytical fitting formulae to N-body simulations) 
dark matter haloes are populated following laws of the kind:\\
$N(M) = N_0(M/M_{\rm min})^{\alpha} \;\;\;
\rm{if} \; M \ge  M_{\rm min}, \nonumber$\\
where N(M) is the number of galaxies within a halo of some mass M, and the parameters $\alpha$ and M$_{\rm min}$ are anti-correlated, in the sense that higher values for $\alpha$ correspond to lower 
values for M$_{\rm min}$. This implies that in the presence of multiple halo occupancy, the values for 
M$_{\rm min}$ found within the HOM scenario will be lower than those obtained via the halo bias model. 
Magliocchetti et al. (2008) found that the difference in the mass estimates coming from the halo bias and HOM approaches 
is of about  0.5 dex both at high ($z\sim 2$) and lower ($z\sim 1$) redshifts.

 In order to homogenize the results, we have then decided to use as a reference model that of Sheth \& Tormen (1999), which provides a working frame that is very similar to that of Mo \& White (1996). Although the HOM approach, which also considers the distribution of galaxies within their dark matter haloes, would in principle be preferable, in practice there are non-trivial complications because it requires a precise knowledge of the behaviour of the two-point correlation function on small scales. It is therefore unapplicable to all those datasets which do not include a large enough number of sources, as is the case of most  high-redshift surveys. The adopted approach is nevertheless valid since: 1) the main aim of our analysis is to compare results coming from different samples so that any possible bias in the determination of $M_{\rm min}$ is not crucial as long as all the mass estimates are evenly 'affected' by it;  and 2) the halo bias model still provides estimates of a physical quantity which is the mass of the parent halo where the galaxies reside as opposed to that of the galactic sub-haloes.

The linear halo bias corresponding to the different sets of sources at the various redshifts can be simply written as:
\begin{eqnarray}
b^{\rm obs}(\bar{z})=\sigma_8/\left[\sigma_8^mD(\bar{z})\right],
\label{eq:bias}
\end{eqnarray}
where $\bar{z}$ is the mean redshift of the considered sample, $D(z)$ is the cosmological growth factor, 
$\sigma_{8}$ is given by eq.~(4) and $\sigma_8^m=0.8$ (see Section 1) is the corresponding local value for the dark matter.  The observed bias can then be compared to its theoretical value, obtained at the same redshift as a function of  $M_{\rm min}$ (Sheth \& Tormen 1999).
Such a comparison returns the values of the minimum masses of the parent haloes hosting the galaxies under exam.

The variation of  $M_{\rm min}$ with cosmic epoch is illustrated in Figure~2, where both symbols and colour coding are as in Figure~1. 
The plot shows that the clear dichotomy seen in the values of the clustering amplitude in the low and in the high redshift regime cannot be accounted for by the redshift evolution of the bias factor at {\bf a} fixed halo mass. While the (small) scatter amongst different measurements within each of the two redshift ranges can be easily explained by different selection techniques, different star formation rates (compare, e.g.  Lin et al. 2012 and Brodwin et al. 2008), and peculiarities in the redshift distribution of the sources (as in van Kampen et al. 2012 who find a peak in the distribution at $z\sim 0.15$ which enhances the amplitude of their correlation function), the much larger discrepancy between $M_{\rm min}$ values at low and high redshifts can only be attributed to different intrinsic properties of the populations of star forming galaxies. 

High-$z$ star-forming galaxies are hosted by very massive, $10^{13}-10^{14}$ $M_\odot$, cluster-like structures, while in the nearby universe even large SFRs (SFR$\simgt 100$ $M_\odot$/yr) are associated to much less massive, $10^{11}-10^{12}$ $M_\odot$, haloes. 

Furthermore, from those works which include highly complete samples (e.g. Magliocchetti \& Maddox 1999, Magliocchetti et al. 2008, 2011, 2013 and Lin et al. 2012), we can estimate the space density of the high-redshift star forming galaxy population. In all cases we find values of a few  $10^{-5}$Mpc$^{-3}$. 
This figure can be compared with the abundance of dark matter haloes with masses greater than the values reported in Table 1  as predicted by the Sheth \& Tormen (1999) mass function. By doing this, we can conclude that the overwhelming majority of dark matter haloes which reach the threshold  mass necessary to host a star-forming galaxy at high redshifts will be inhabited by (at least) one of such objects. In other words, massive star-forming galaxies at redshifts around 2 are a widespread event, 
their observed paucity  being simply due to the paucity of high mass dark matter haloes at the considered redshifts. Since the above samples contain sources selected in redshift ranges $\Delta z\simeq 0.7$, this implies a life-time for this intense star-forming phase at $z\sim 2$ of about 1 Gyr.

On the other hand, in the local Universe only a fraction of (the much smaller and therefore much more numerous)  virialized haloes will host a powerful star-forming event, and this fraction is found  to decrease to lower redshifts (cfr Magliocchetti et al. 2011).

If we assume a  linear downsizing trend, we can parametrize the mass dependence  of the star-forming galaxy population on look-back time as ${\rm log_{10}}$[$M_{\rm min} /M_\odot]=\alpha \cdot z+\beta$, with $\alpha=1.0^{+0.05}_{-0.10}$ and $\beta=11.35^{+0.20}_{-0.10}$. The fit is reproduced in Figure 2 by the dashed line.

\section{The fate of $z\sim 2$ star-forming galaxies}

\begin{figure}
\begin{center}
\includegraphics[scale=0.45]{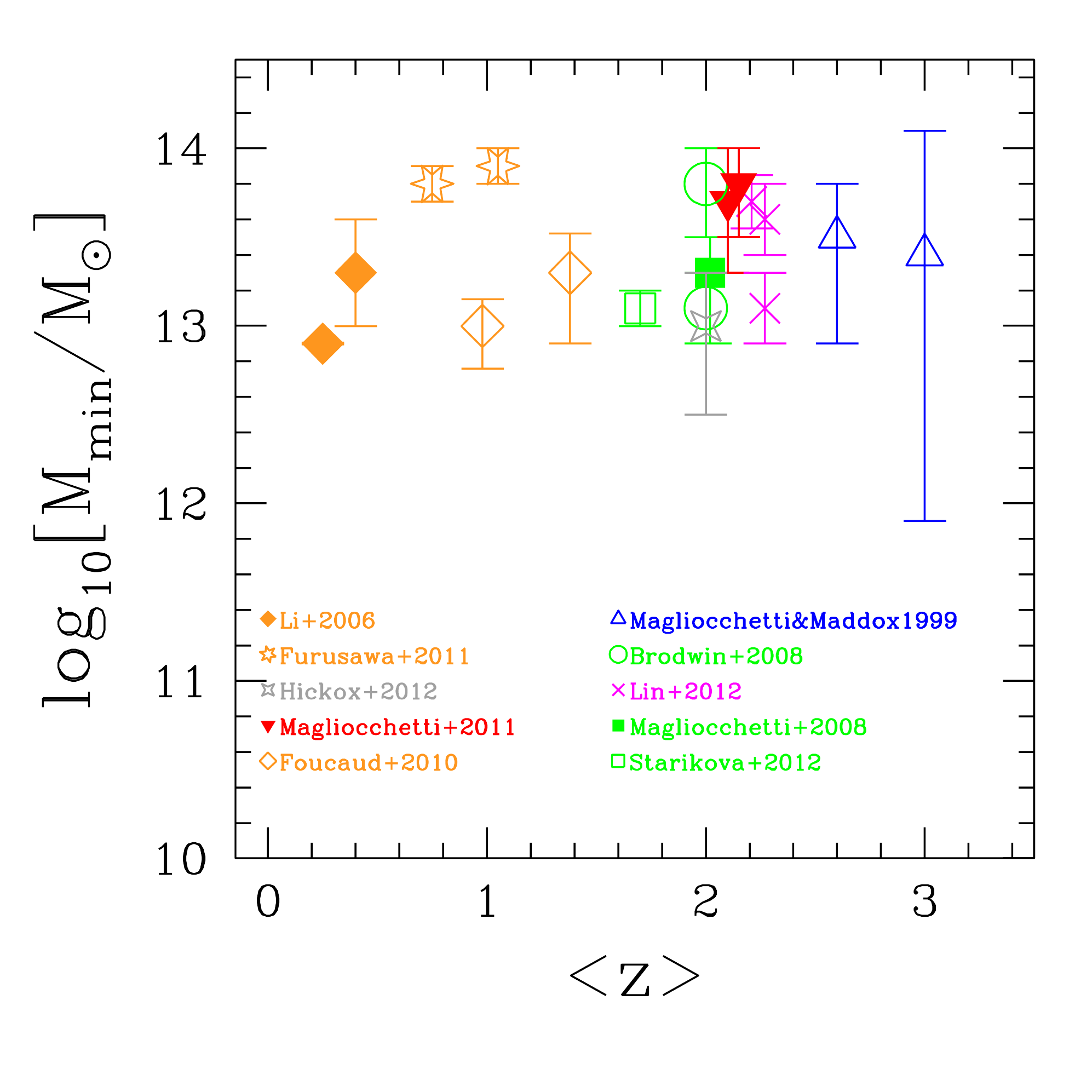}
\caption{The clustering properties of high-$z$ galaxies with intense star-formation (plotted for $z\simgt 1.7$) match those of passively evolving $z \simlt 1.5$ galaxies with stellar masses in the range $\simeq 10^{11}$--$10^{12}\,M_\odot$. Symbols and colour coding for star-forming galaxies are the same as in Figure 2.
\label{fig:mass}}
\end{center}
\end{figure}

The analysis performed so far has clearly shown that high-$z$ massive star-forming galaxies have no low-$z$ counterpart. It is then natural to ask how and into which objects they evolve. Other works (e.g., amongst the many, Somerville et al. 2004, Brodwin et al. 2008, Hickox et al. 2012) have investigated the cosmic evolution of a specific galaxy population via comparison of clustering measurements at low and high redshifts. Our approach is however different, since we will not investigate such an evolution in a "blind way ", i.e. by extrapolating halo masses at lower redshifts and looking for matching values in the galaxy zoo. What we have instead is a prediction which we will test with the use of the available data. 

In fact, according to the physical model by Granato et al. (2004), further developed by Lapi et al. (2006; 2011), a high-$z$ galaxy with a $\hbox{SFR}\sim 300 M_\odot$/yr (typical for the objects considered in this work) spends about 0.5--1~Gyr in this intense star-forming phase. This value is in agreement with that obtained in Section 4 from investigation of the number densities of such objects. At the end of this stage, the galaxy will have formed $M_* \sim 1.5\hbox{--}3\cdot 10^{11} M_\odot$ stellar masses and will rapidly evolve into a passive, red and dead, source.

To check whether this view is consistent with the {\bf available} clustering data, we have compared our estimates of the minimum halo masses for high-$z$ star-forming galaxies with those obtained for passively evolving galaxies at $z<1.5$ endowed with stellar masses in the range $\sim [10^{11}-10^{12}] M_\odot$. Specifically, at low redshift ($z\sim 0.3$) we consider the clustering analysis by Li et al. (2006) performed on SDSS galaxies with  stellar masses respectively ranging between $10^{11} \le M_*/M_\odot <10^{11.5}$ and $10^{11.5} \le M_*/M_\odot <10^{12}$. At  higher redshifts ($z\sim 0.6-1.3$) we consider the results by Foucaud et al. (2010) and Furusawa et al. (2011), respectively for $K$-selected galaxies with $10^{11} \le M_*/M_\odot  \le 10^{12}$ taken from the Palomar Observatory Wide-field Infrared Survey  and for $K$-selected galaxies with $10^{10.86} \le M_*/M_\odot  <10^{11.26}$ from the Subaru/XMM-Newton Deep Survey + UKIDSS UDS fields.

Halo masses obtained with a Mo \& White (1996) formalism were directly provided by the authors in the two latter cases,  while Li et al. (2006) provide estimates for the bias factor $b$, which can be easily converted into minimum halo masses by following the approach described in Section 4. The results are presented in Fig.~3. Filled diamonds are for the Li et al. (2006) sample, the lower point representing the lower stellar mass interval, while open stars are for the results by Furusawa et al. (2011) and open diamonds for Foucaud et al. (2010).

In spite of some scatter, minimum halo masses between $10^{13}$ and $10^{14}$ M$_\odot$ are found for all the samples. This range of values  perfectly mirrors those derived for  the $z\simgt 1.5$, actively star-forming galaxies  previously considered in this work.  We note that,  although halo masses are expected to increase with time, for haloes of $\sim 10^{13}
M_\odot$, the evolution between z=2 and z=0 is relatively modest (0.5 dex or so; e.g. Fakhouri et al. 2010), and thus comparable or smaller than the scatter between the points given in Figure 3.

The above similarity strongly points to an evolutionary link between vigorous star-forming galaxies at $z\simgt 1.5$ and $z\simlt 1.5$ passively-evolving sources with a large stellar mass content.

\section{Conclusions}

We have highlighted a striking dichotomy in the clustering amplitude of  very active star-forming galaxies (SFRs ranging from $\sim 20-30$ $M_\odot$/yr up to a few $\times 10^3$ $M_\odot$/yr) at low and high redshifts ($z \simlt 1$ and $z \simgt 2$). The comoving clustering radii, estimated in a uniform way, are approximately constant within each redshift range but jump by about a factor of 3 from $z \simlt 1$ (where they are 4--6 Mpc) to $z \simgt 2$ (where they range from $\sim 11$ to $\sim 20$ Mpc). We argue that this leap cannot be due to some selection bias or to the luminosity dependence of the clustering amplitude. 

The variation of the clustering signal is reflected in the distribution of halo masses. Regardless of the chosen selection criterion, {\bf all} high-$z$ galaxies undergoing an intense process of star formation are hosted by very massive, $10^{13}-10^{14}$ $M_\odot$, cluster-like structures, while in the nearby universe even large SFRs ($SFR \simgt 100$ $M_\odot$/yr) are associated to much smaller, $10^{11}-10^{12}$ $M_\odot$, haloes. This implies that the actively star forming galaxy population observed at $z \simgt 2$ is not the same we see at $z \simlt 1$.
 We argue that  the lack of massive star-forming galaxies in the $z\simlt 1.5$ universe is due to a rather rapid (characteristic timescale of the intense $z\sim 2$ star-forming phase $\sim 1$~Gyr) evolution  of the active $z\simgt 2$ population into passive sources with very large, $M_*\simeq [10^{11} - 10^{12}] M_\odot$, stellar masses.

We stress that the striking dichotomy observed in the physical properties of low-z and high-z galaxies undergoing intense star formation is confirmed at all wavelengths capable of probing the star formation regime. It is worth noticing though that high redshift galaxies with low-to-moderate star formation activity (SFR$\simlt 20$ M$_\odot$/yr) show a clustering behaviour which is different from that of their more active companions. In fact, these sources exhibit much smaller clustering lengths (e.g. Adelberger et al. 2005; Xia et al. 2012;  Lin et al. 2012; Viero et al. 2013), comparable with those of their lower-redshift counterparts. This issue will be investigated in a forthcoming paper.

 The above findings strongly argue against merging as the main trigger of the highly enhanced star-forming phase of galaxies at $z\sim 2$. In fact, scenarios 
 which envisage modest-sized merger-induced starbursts fail at reproducing the very large halo masses estimated from all the available clustering data. On the other hand, the observed space densities and inferred life-times of the star-burst phase disagree with those predicted by models which advocate mergers of large units as the primary driver of the phenomenon. 
 
 The gap in the clustering data between $z\simeq 1$ and $z\simeq 2$ does not allow us to determine yet how the transition from the high-mass/high-redshift  to the low-mass/low-redshift star-forming regime occurs: is it gradual or abrupt? Key information on this point will come from the large and deep surveys that will be provided by Euclid and SKA.\\

 \noindent
{\bf Acknowledgements}
We wish to warmly thank the anonymous referee for the constructive comments which greatly helped at improving the paper.

\end{document}